\documentclass[12pt]{spieman}  % 12pt font required by SPIE;
\usepackage{amsmath,amsfonts,amssymb}
\usepackage{graphicx}
\usepackage{setspace}
\usepackage{tocloft}
\usepackage[super]{nth}
\usepackage{lineno}
\usepackage{xcolor}
\usepackage{soul}

\usepackage{ulem}
\usepackage{marginnote}

\title{Three-sided Pyramid Wavefront Sensor. II.  Preliminary Demonstration on the new CACTI Testbed}

\author[a]{Lauren Schatz}
\author[b]{Johanan Codona}
\author[c]{Joseph D. Long}
\author[c]{Jared R. Males}
\author[b]{Weslin Pullen}
\author[a]{Jennifer Lumbres}
\author[c]{Kyle Van Gorkom}
\author[d]{Vincent Chambouleyron}
\author[c]{Laird M. Close}
\author[d]{Carlos Correia}
\author[e]{Olivier Fauvarque}
\author[f,d]{Thierry Fusco}
\author[a,c,h,i]{Olivier Guyon}
\author[b]{Michael Hart}
\author[d]{Pierre Janin-Potiron}
\author[g]{Robert Johnson}
\author[j]{Nemanja Jovanovic}
\author[g]{Mala Mateen}
\author[f,d]{Jean-François Sauvage}
\author[d]{Benoit Neichel}

\affil[a]{University of Arizona, Wyant College of Optical Sciences, Tucson, Arizona, United States}
\affil[b]{Hart Scientific Consulting International LLC, Tucson, Arizona, United States}
\affil[c]{University of Arizona, Steward Observatory, Tucson, Arizona, United States}
\affil[d]{Aix Marseille Univ, CNRS, CNES, LAM, Marseille, France}
\affil[e]{IFREMER, Laboratoire Detection, Capteurs et Mesures (LDCM), Centre Bretagne, ZI de la Pointe du Diable, CS 10070, 29280, Plouzane, France}
\affil[f]{DOTA, ONERA, Université Paris Saclay, Palaiseau, France}
\affil[g]{Air Force Research Lab, Starfire Optical Range, Kirtland Air Force Base, New Mexico, United States}
\affil[h]{Subaru Telescope, National Astronomical Observatory of Japan, National Institutes of Natural Sciences (NINS), 650 North A‘ohōkū Place, Hilo, Hawaii 96720, USA}
\affil[i] {Astrobiology Center of NINS, 2-21-1, Osawa, Mitaka, Tokyo, 181-8588, Japan}
\affil[j]{Caltech Optical Observatories, California Institute of Technology, Pasadena, CA 91125}

\cftpagenumbersoff{figure}
\cftpagenumbersoff{table} 
\begin{document} 
\maketitle

\begin{abstract}

The next generation of giant ground and space telescopes will have the light-collecting power to detect and characterize potentially habitable terrestrial exoplanets using high-contrast imaging for the first time. This will only be achievable if the performance of Giant Segmented Mirror Telescopes (GSMTs) extreme adaptive optics (ExAO) systems are optimized to their full potential. A key component of an ExAO system is the wavefront sensor (WFS), which measures aberrations from atmospheric turbulence. A common choice in current and next-generation instruments is the pyramid wavefront sensor (PWFS). ExAO systems require high spatial and temporal sampling of wavefronts to optimize performance, and as a result, require large detectors for the WFS. We present a closed-loop testbed demonstration of a three-sided pyramid wavefront sensor (3PWFS) as an alternative to the conventional four-sided pyramid wavefront (4PWFS) sensor for GSMT-ExAO applications on the new Comprehensive Adaptive Optics and Coronagraph Test Instrument (CACTI). The 3PWFS is less sensitive to read noise than the 4PWFS because it uses fewer detector pixels. The 3PWFS has further benefits: a high-quality three-sided pyramid optic is easier to manufacture than a four-sided pyramid. We detail the design of the two components of the CACTI system, the adaptive optics simulator and the PWFS testbed that includes both a 3PWFS and 4PWFS. We detail the error budget of the CACTI system, review its operation and calibration procedures, and discuss its current status. A preliminary experiment was performed on CACTI to study the performance of the 3PWFS to the 4PWFS in varying strengths of turbulence using both the Raw Intensity and Slopes Map signal processing methods. This experiment was repeated for a modulation radius of 1.6 $\lambda/D$ and 3.25 $\lambda/D$. We found that the performance of the two wavefront sensors is comparable if modal loop gains are tuned.

\end{abstract}

% Include a list of up to six keywords after the abstract

% Include email contact information for corresponding author
{\noindent \footnotesize\textbf{*}Lauren Schatz,  \linkable{laurenhschatz@gmail.com} }

\begin{spacing}{1.25}   % use double spacing for rest of manuscript

\section{Introduction}
\label{sect:intro}  % \label{} allows reference to this section

High-contrast imaging refers to a collection of techniques used to image spatially resolved faint objects next to bright sources. Targets include but are not limited to circumstellar disks \cite{rodigas2014morphology}, active galactic nuclei \cite{imanishi2020subaru}, and exoplanets\cite{bowler2016imaging}. Exoplanets have flux ratios of 10$^{-4}$ or deeper contrast with respect to their host stars. Terrestrial exoplanets are typically fainter and have close orbits to the host star and therefore require even greater contrasts for detection, 10$^{-8}$ or deeper. These high-contrast ratios present challenges in directly imaging such faint objects. Overcoming this contrast problem requires a two-fold solution. The starlight needs to be suppressed by a coronagraph, and the resulting high-contrast region, called the dark-hole, needs to be maintained throughout the observation through extreme adaptive optics (ExAO) and wavefront sensing and control (WFS$\&$C) techniques. ExAO systems operate by directing light from a guide star to a wavefront sensor, which measures the phase error of the starlight wavefront. A computer then sends commands to shape a deformable mirror (DM) to correct for the phase error, forming a closed feedback loop that compensates for most of the atmospheric distortion. The corrected beam is then passed to a coronagraph which blocks the light from the on-axis star while allowing off-axis light from the target of interest to pass.
% These speckles evolve with time\cite{goebel2016evolutionary} so further methods of WF$\&$C in addition to the ExAO closed loop is needed to null speckles \jrmadd{$\leftarrow$ correct, but relevant to this paper?}.  

% GET RID OF DESCRIPTIONS START LISTING, CURRENT TESTBEDS INCLUDE...
%  \jrmcom{How is coronagraphy relevant?  This ought to be in intro of dissertation overall, but not in this paper} \jrmrmv{One area of research is the development of different coronagraphs for telescopes that have complex apertures. There are also test-beds focused on advancing ExAO and WF$\&$C. This included techniques such as Linear Dark Field Control (LDF), Electric Field Conjugation (EFC), and Low Order Wavefront Sensing (LOWFS).}
The next generation of giant ground and space telescopes will have the light-collecting power to detect and characterize potentially habitable terrestrial exoplanets using high-contrast imaging for the first time. Smaller, terrestrial planets that orbit close to their stars require larger contrasts than larger exoplanets that have broad orbits \cite{males2014direct}. This will only be achievable if the performance of Giant Segmented Mirror Telescopes (GSMT) ExAO systems are optimized to their full potential. The ground-based GSMTs include the Thirty Meter Telescope (TMT) \cite{chisholm2020thirty}, the Giant Magellan Telescope (GMT) \cite{fanson2020overview}, and the European Extremely Large Telescope (E-ELT)\cite{ramsay2020eso}. Various testbeds are advancing technology and techniques to enable exoplanet imaging on the next generation of telescopes. Current testbeds include the High Contrast Imager for Complex Aperture Telescopes (HiCAT)\cite{2014SPIE.9143E..27N} at the Space Telescope Science Institute and the LAM-ONERA On-sky Pyramid Sensor (LOOPS)\cite{janin2019adaptive}. There are many other examples of testbeds \cite{janin2019adaptive} \cite{jovanovic2018high} \cite{males2020magao} \cite{jovanovic2015subaru} and many are listed on the Community of Adaptive Optics and High Contrast testbeds website\cite{CHAOTIC}. A recent summary of current coronagraphy testbeds for space missions can be found in the decadal white paper by Mazoyer, et al.\cite{mazoyer2019high}
%\jrmcom{this is a great summary}

%\jrmcom{Moving this up, and asking for a little more}
A key component of an ExAO system is the WFS, which measures aberrations from atmospheric turbulence. A common choice in current and next-generation instruments is the pyramid wavefront sensor (PWFS)\cite{ragazzoni2002pyramid}. The PWFS is a highly sensitive wavefront sensor able to measure wavefront errors at high speeds. The sensitivity and linear range of the PWFS can be tuned by dynamic modulation, making the PWFS robust to different seeing conditions.  

\begin{figure}
\centering
\includegraphics[width=0.5\textwidth]{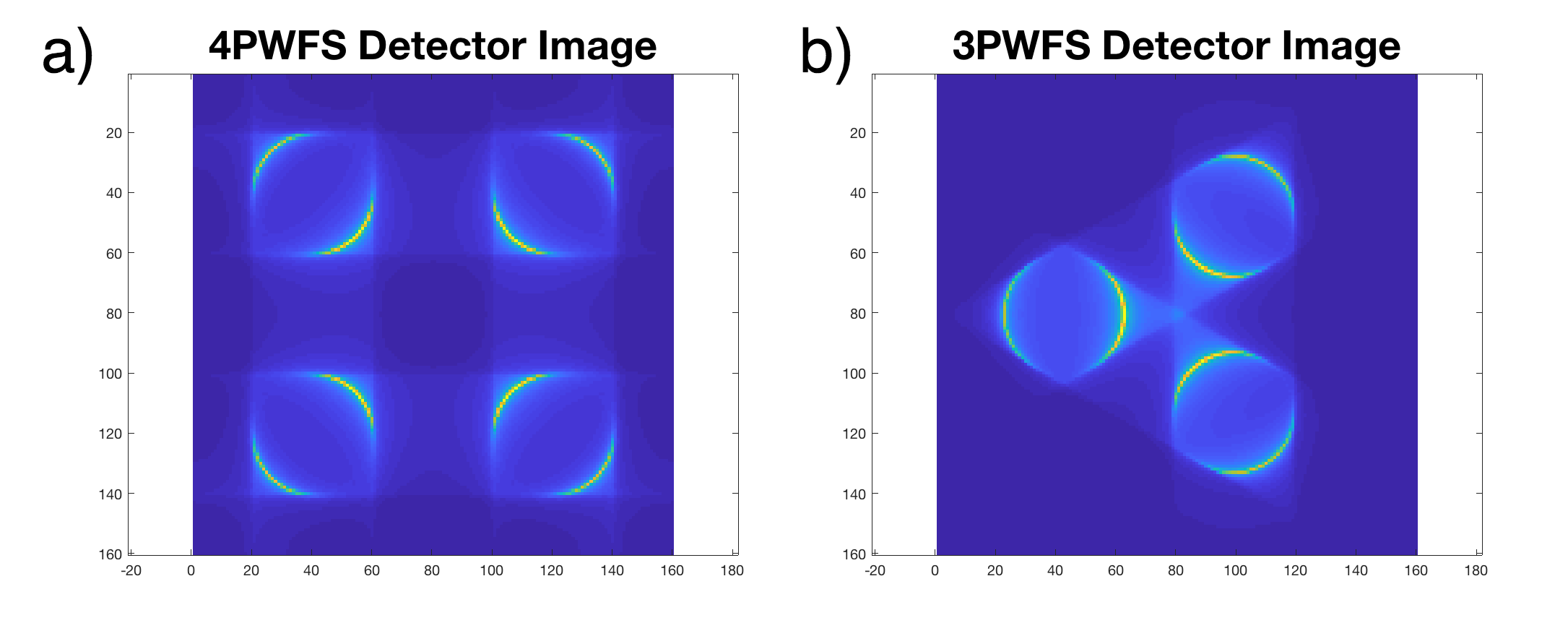}
\caption{Simulated PWFS detector images under no modulation for the 4PWFS (a) and the 3PWFS (b).}
\label{fig:PWFSpupils}
\end{figure}

The PWFS performs a Foucault test in two dimensions. Light from the telescope is focused onto a glass pyramid tip where it is split and then the pupil plane is re-imaged onto a detector. The result is copies of the telescope pupil that contain intensity fluctuations related to the wavefront phase. All PWFS currently on telescopes use a four-sided pyramid (4PWFS), resulting in four pupil images. A three-sided pyramid wavefront sensor (3PWFS) forms three pupil images instead resulting in fewer detector pixels needed for wavefront sensing. Figure \ref{fig:PWFSpupils} shows simulated detector images for the 4PWFS and the 3PWFS under zero modulation. In low light conditions where read noise dominates this translates in a signal-to-noise ratio gain for $\sqrt{\frac{4}{3}}$ for measurements taken by a 3PWFS. The 3PWFS has further benefits: a high-quality three-sided pyramid optic is easier to manufacture than a four-sided pyramid. Schatz et al.\cite{schatz2021three} determined in simulation that the 3PWFS meets and can exceed the performance of a 4PWFS for a detector with high read noise. 

%ExAO systems require high spatial and temporal sampling of wavefronts to optimize performance and as a result, require large detectors.

Here we present the Comprehensive Adaptive Optics and Coronagraph Test Instrument (CACTI), which was designed with the flexibility to support visiting instruments and to be easily re-configurable to perform multiple experiments.  We first describe the design of CACTI, review its operation and calibration procedures, and discuss its current status. We then discuss a preliminary experiment performed on CACTI with a visiting 3PWFS to explore an alternative wavefront sensor architecture for GSMT-ExAO. Both a 3PWFS and 4PWFS were integrated into  CACTI for an initial demonstration of the operation of the 3PWFS. As an incremental step in developing the 3PWFS, we performed a preliminary study comparing the performance of the 3PWFS and the 4PWFS under different turbulence strengths. The goal of these tests was to show that the 3PWFS can reconstruct a wavefront with an accuracy comparable to the 4PWFS, not to provide a definitive measurement of the maximum Strehl producible by a PWFS achievable in on-sky conditions. Finally, we discuss the outcome of these experiments.

\section{Design of CACTI}
CACTI was designed to model a full end-to-end adaptive optics system with the flexibility to support multiple experiments. In the configuration described here, CACTI consisted of two components: an adaptive optics simulator and a pyramid wavefront sensor testbed. In the following sections, we describe the optical design in detail.

\subsection{Adaptive Optics Simulator}

\begin{figure}
    \centering
    \includegraphics[width=0.8\textwidth]{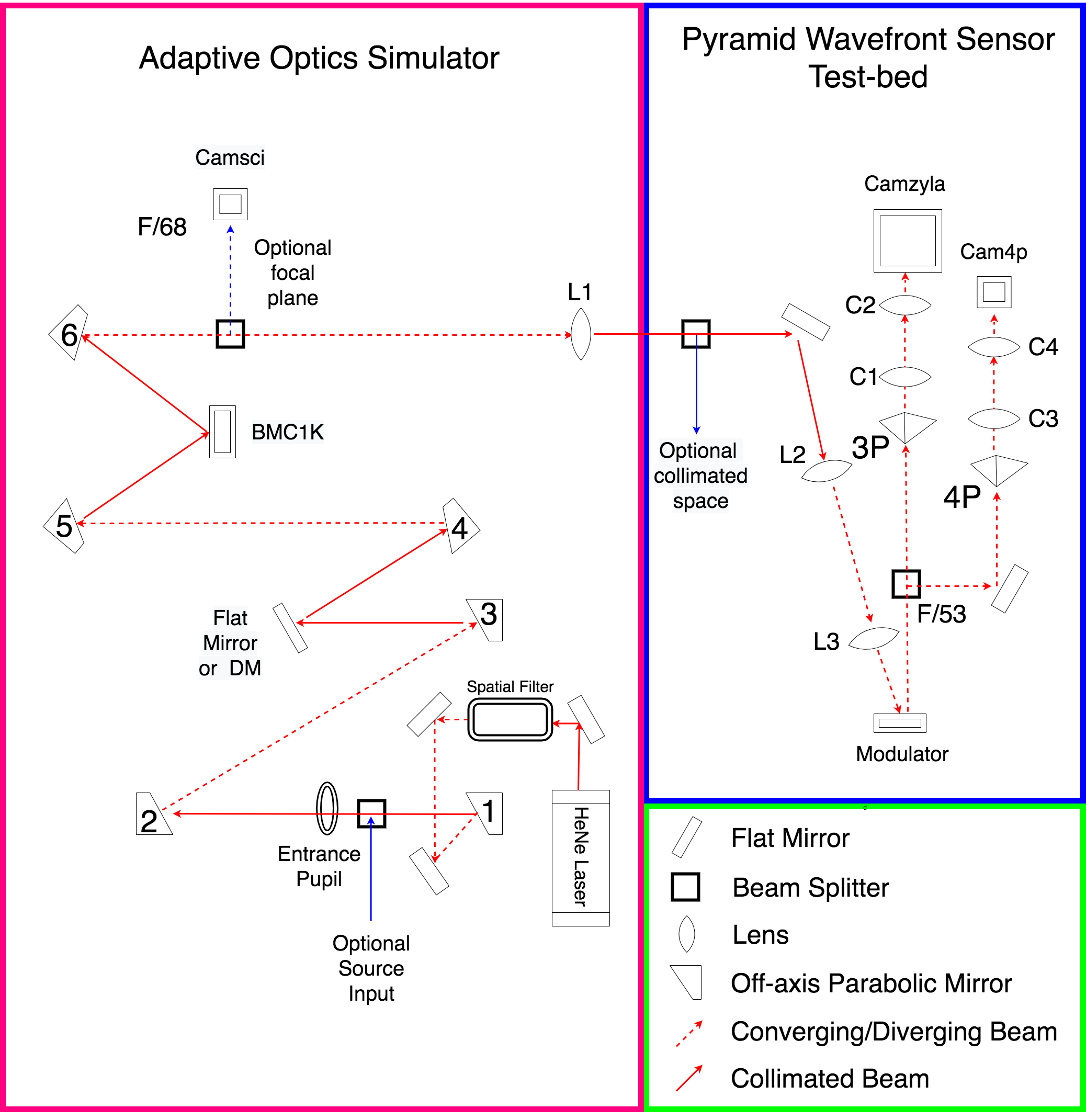}
    \caption{In the current configuration CACTI consists of an Adaptive Optics Simulator and the PWFS testbed. Beamsplitters are used in CACTI to provide access to focal planes, collimated spaces, and additional sources. In the current configuration light is relayed through the AO simulator onto the 1024 actuator Boston Micromachine 1K (BMC1K) DM OAP mirrors. The light is then passed to the PWFS testbed which includes a modulation mirror, a 4PWFS and a 3PWFS. The F-number on the pyramid tip is F/53, and the F-number on camsci is 68.}
    \label{fig:CACTI}
\end{figure}

CACTI was designed to simulate atmospheric turbulence in a complete closed-loop AO system. Table \ref{tab:CACTItable} summarizes the main components. The layout of the optical table is shown in Figure \ref{fig:CACTI}. A Helium-Neon (HeNe) laser (wavelength 633 nm) is used for initial alignment and testing.  The light is passed to a spatial filter with a 10 $\mu$m pinhole to clean up any wavefront errors from our laser source and ensure that the start of the system is an unresolved point source. The point source is then collimated by an off-axis parabolic (OAP) mirror to simulate starlight coming from infinity. There are six 2-inch diameter OAP mirrors in total that form the pupil relays of the system. All are cored from the same parent with $\lambda /10$ peak-to-valley surface quality, where the wavelength, $\lambda$, is 633 nm. Each OAP has a focal length of 375.25 mm and an off-axis angle of 23 degrees.

 A 50/50 beamsplitter cube is placed into the collimated beam after the first OAP as an optional input for another collimated light source. After the beamsplitter, a 7.5 mm diameter circular clear aperture mask is placed to define the entrance pupil of CACTI. The first pupil relay formed by the \nth{2} and \nth{3} OAP mirrors re-images the entrance pupil onto a flat mirror that is mounted on a kinematic base. The flat mirror is intended to be removed and replaced by a DM in the future. The second pupil relay created by the \nth{4} and \nth{5} OAP mirrors relays the pupil onto a 1024 actuator deformable mirror (BMC1K) with 300 $\mu m$ pitch. In the experiments described here, the DM is used to simulate the atmosphere, correct the errors in closed-loop with the wavefront sensor, and correct for common path errors from misalignments. The last OAP focuses the light to the final focal plane in the AO simulator. A 50/50 beamsplitter is placed in this converging beam so that an additional focal plane can be accessed. In the current configuration of CACTI, we have placed a CMOS camera as our science camera (Camsci) at this focal plane. The F-number at our science focal plane is F/68. A ND filter of 2.5 was inserted into the system to prevent over-saturation on the cameras. 

\begin{table}
	\begin{center}
		\begin{tabular}{ | l| l | }
			\hline
			\textbf{Component}& \textbf{Description}\\ \hline
			Light source & Helium-Neon laser 633 nm $\lambda$\\ \hline
			Spatial filter & 10 $\mu$m Pinhole \\ \hline
			Off-axis parabolic mirrors & 375.25 mm Focal length, 23$^{\circ}$ off-axis angle \\ \hline
			Entrance pupil & 7.5 mm Clear aperture mask \\ \hline
			Deformable mirror & Boston Micromachines, 1024 Actuators, 9.6 mm in diameter \\ \hline
			Beamsplitters & 50/50 Cube beamsplitter \\ \hline
			Modulation mirror &  PI S-331 Piezo-actuator stage \\ \hline
			3PWFS Optic & Fused silica glass monolith \\ \hline
			4PWFS Optic & Crossed roof prisms \\ \hline
			Science camera (Camsci) & Basler ACE acA640-750um CMOS\\ \hline
			3PWFS camera (Camzyla) & Zyla 4.2+ sCMOS detector \\ \hline
			4PWFS camera (Cam4p) & Basler ace acA720-520um CMOS \\ \hline
			Lens 1 (L1) & Doublet 500 mm focal length  \\ \hline
			Lens 2 (L2) & Custom doublet lens  \\ \hline
			Lens 3 (L3) & Custom achromatic air-spaced triplet  \\ \hline
			3PWFS Camera lens 1 (C1) & Doublet 30 mm focal length \\ \hline
			3PWFS Camera lens 2 (C2) & Doublet 30 mm focal length \\ \hline
			4PWFS Camera lens 1 (C3) & Doublet 50 mm focal length \\ \hline
			4PWFS Camera lens 2 (C4) & Doublet 30 mm focal length \\ \hline
			\textbf{AO Parameter}& \textbf{Description}\\ \hline
			Number of illuminated actuators &  545 \\ \hline
			Loop Speed & 400 Hz  \\ \hline
			WFS Camera FPS & 400 Hz  \\ \hline
			WFS Camera exposure time & 0.001 sec  \\ \hline
			WFS Modulation speed & 1 kHz \\ \hline
			
		\end{tabular}
	\end{center}
	\caption{Descriptions of the components in CACTI.}
	\label{tab:CACTItable}
\end{table}

\subsection{System Characterization}\label{characterization}

\begin{figure}
 	\centering
 	\includegraphics[width=1\textwidth]{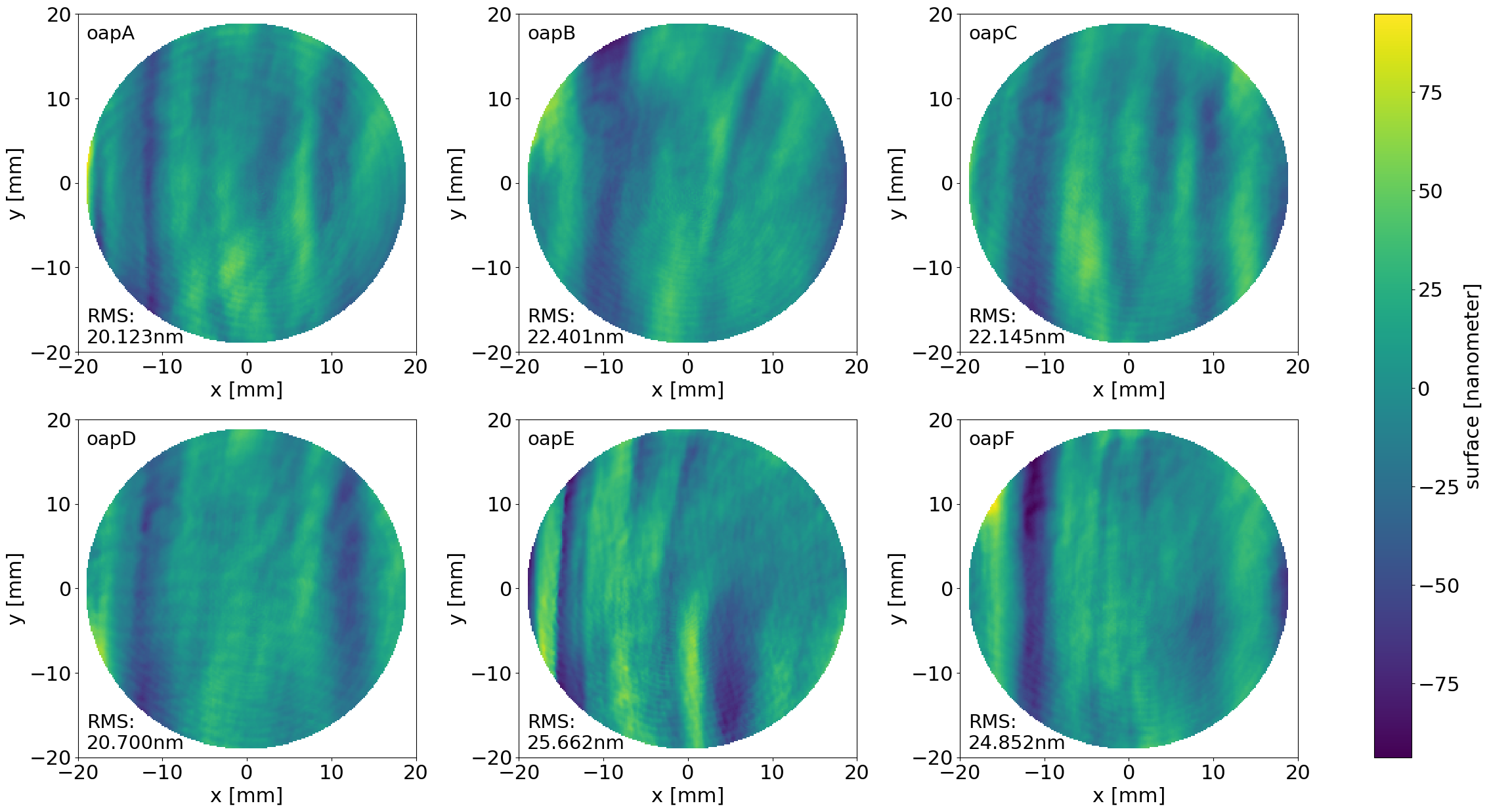}
 	\caption{Surface measurements of the OAP mirrors in CACTI. In color scale is the surface error in nanometers. All mirrors were cored from the same parent and specified at $\lambda/10$ surface quality.  }
 	\label{fig:OAPfig}
 \end{figure}

\begin{figure}
  	\centering
  	\includegraphics[width=0.8\textwidth]{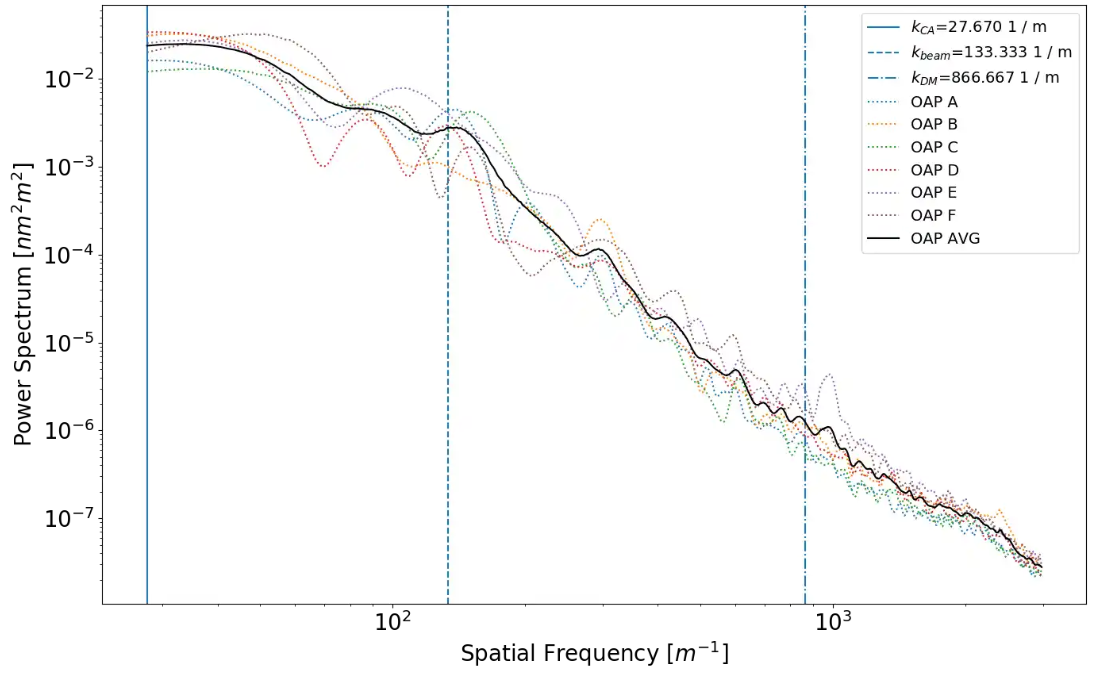}
  	\caption{Power spectrum estimate versus spatial frequency of the OAP mirrors. The solid black line is the average PSD estimate for all OAPs. Overlaid on this plot are vertical blue lines representing the lower spatial frequency boundary used to calculate the root-mean-square (RMS) surface error for different parameters such as beam size and DM correction. These boundaries can be read as all spatial frequencies to the right of each line are included in the RMS calculation. The first line ($k_{CA}$) represents the 75$\%$ clear aperture of the OAP mirror. The middle line ($k_{beam}$) represents the lower spatial frequency limit from our 7.5 mm beam diameter. The final line ($k_{DM}$) represents the lower spatial frequency limit corrected by the DM for the 7.5 mm beam diameter in our PSF sharpening process.}
  	\label{fig:PSD}
  \end{figure}

 A thorough characterization of the key optical components in CACTI was performed to determine the AO system performance using a power spectral density (PSD) analysis of surface measurements developed by Lumbres (2021)\cite{jhendissertation}.The interferometric surface measurements of the six OAP mirrors were provided by the manufacturer. Post-processing was done on the measurements to mask the aperture to 75$\%$ clear aperture and to remove piston, tip and tilt. The final processed images of the OAP surfaces are displayed in Figure \ref{fig:OAPfig}. Shown in the color scale is the surface height error in nanometers. In this analysis, the surface maps were used to generate a 2D plot of power spectrum [$nm^2 m^2$] versus spatial frequency [$m^{-1}$] for each OAP mirror. The plot of the power spectrum estimate of each OAP mirror and the calculated average power spectrum is shown in Figure \ref{fig:PSD}. Overlaid on this plot are vertical blue lines representing the lower spatial frequency boundary used to calculate the root-mean-square (RMS) surface error for different parameters such as beam size and DM correction. These boundaries can be read as all spatial frequencies to the right of each line are included in the RMS calculation. The first line is the spatial frequency boundary for the  38.1 mm clear aperture of the system. In CACTI the OAPs are illuminated by a beam diameter of 7.5mm and this spatial frequency lower bound is represented by the $k_{beam}$ line. Additionally in CACTI we sharpen the PSF using a grid-search procedure we call the eye-doctor\cite{van2021characterization} \cite{bailey2014large}. Using the DM we correct up to the \nth{35} Zernike mode, which corresponds to a correction of about 6 cycles per aperture. The line best representing the spatial frequencies present in the CACTI system is therefore the $k_{DM}$ line which reflects the correction by the DM for the 7.5 mm beam diameter.

\begin{figure}
    \centering
    \includegraphics[width=1\textwidth]{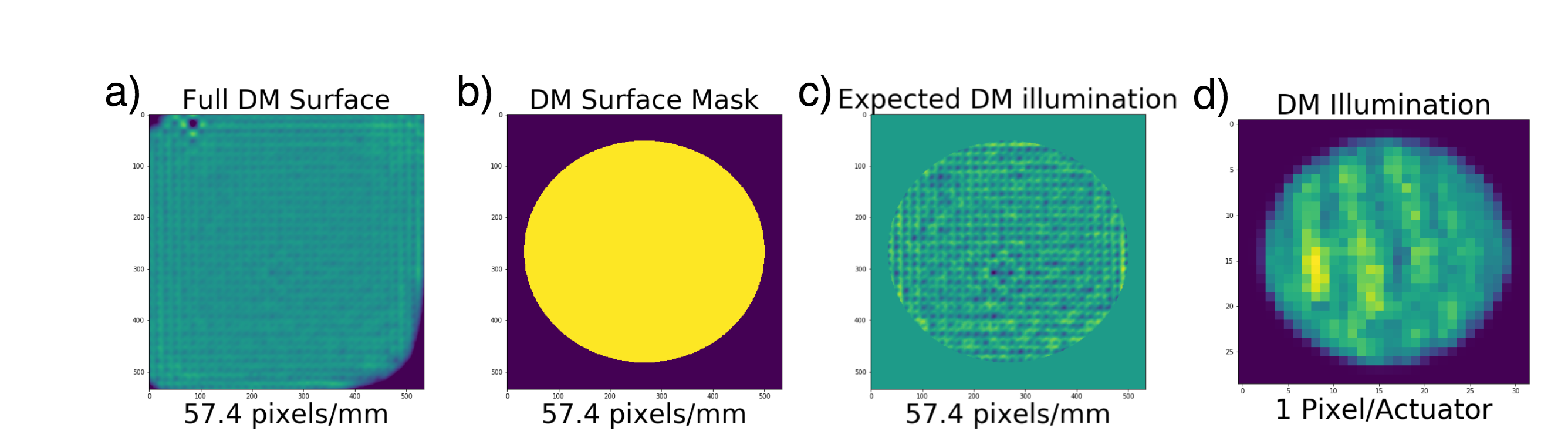}
    \caption{A. Surface map of the flattened DM measured by an interferometer. B. The estimated DM surface illumination pattern by our optical design. C. The surface map of the flattened DM masked by the DM surface mask given in B). D. The calculated DM illumination pattern from our calibration procedure.}
    \label{fig:dmMaps}
\end{figure}

The deformable mirror was flattened using an interferometer by measuring an poke interaction matrix of the DM and interferometer and iteratively flattening the surface in closed loop. Figure \ref{fig:dmMaps}.A shows the full measured DM surface. Figure \ref{fig:dmMaps}.B is the mask representing the expected illumination pattern on our DM calculated by our optical design which predicted an actuator illumination of 27.2 in the X-direction, and 25 actuators in the Y-direction. This mask is projected onto our measured DM surface in Figure \ref{fig:dmMaps}.C. Figure \ref{fig:dmMaps}.D is the measured DM illumination calculated by our control software during system calibration. There are 28 actuators illuminated in the X-direction, and 26 in the Y-direction. Van Gorkom et al. (2021)\cite{van2021characterization} details the characterization of the influence functions for modal control of the 1024 DM. The RMS surface error of the DM is 11.2 nm across the full surface, and 3 nm across our beam diameter. 

In CACTI we use a grid search to apply different amplitudes of Zernike modes on the deformable mirror. The algorithm saves the combination of modes as a single DM set point that maximizes the Strehl Ratio at the focal plane where the algorithm is implemented. We, therefore, assume that the static aberrations in the CACTI system caused by surface errors of the optics or misalignment are corrected by the deformable mirror up to the spatial frequency of the modes used by the eye-doctor. On CACTI we correct up to the first 35 Zernike polynomials which correspond to correcting about 6 cycles per aperture in spatial frequency. Using the estimated PSDs plotted in Figure \ref{fig:PSD} we integrate the power spectrum to determine the RMS surface error of each OAP over the beam diameter, factoring both the DM correction and the angle of incidence of the beam on the OAP surface. Table \ref{tab:OAPtable}  summarizes the RMS values for the expected RMS optical path difference (OPD) caused by these errors, where we define the OPD RMS error as twice the RMS surface error for a mirror. Included in Table \ref{tab:OAPtable} are the estimated RMS surface values for our flattened DM as well. For the combined wavefront error from the CACTI OAPs we estimate an OPD RMS error of 26.917 nm over our 7.5 mm beam diameter. Using the Maréchal approximation we calculate the highest possible Strehl Ratio for the CACTI adaptive optics simulator to be 0.93 (93 $\%$) after correction and only considering the fitting error from optical surfaces.

% The other optics in the CACTI system are off the shelf lenses quoted at a surface quality of $\lambda/4$ PV which translates to an RMS surface of 39.56 nm following the convention set by Schwertz\cite{schwertz2010useful} at a wavelength of 633 nm. 

\begin{table}
	\begin{center}
		\begin{tabular}{ | l| l |}
			\hline
			\textbf{Component}&  \textbf{OPD RMS Surface ($k_{DM}$)}\\ \hline
			OAP A & 3.168 nm  \\ \hline
			OAP B & 4.021 nm  \\ \hline
			OAP C & 3.236 nm  \\ \hline
			OAP D & 3.672 nm \\ \hline
			OAP E & 5.428 nm \\ \hline
			OAP F & 4.392 nm \\ \hline
			Average OAP & 4.001 nm \\ \hline
			DM  & 3 nm \\ \hline
			\end{tabular}
		\end{center}
	\caption{RMS OPD error contributed to the CACTI error budget by each OAP after correction by the DM through PSF sharpening. RMS OPD error listed is over the 7.5mm beam diameter and accounting for angle of incidence.}
	\label{tab:OAPtable}
\end{table}

%\begin{table}
%	\begin{center}
%		\begin{tabular}{ | l| l |l|l |}
%			\hline
%			\textbf{Component}& \textbf{RMS Surface ($k_{CA}$)}& \textbf{RMS Surface ($k_{beam}$)}& \textbf{RMS Surface ($k_{DM}$)}\\ \hline
%			OAP A & 19. 399 nm & 11.570 nm & 1.019 nm \\ \hline
%			OAP B & 21.405 nm & 9.805 nm & 1.301 nm \\ \hline
%			OAP C & 21.625 nm & 14.221 nm & 1.058 nm \\ \hline
%			OAP D & 18.922 nm & 8.955 nm & 1.107 nm \\ \hline
%			OAP E & 24.995 nm & 13.908 nm & 1.264 nm \\ \hline
%			OAP F & 24.198 nm & 10.015 nm & 1.264 nm \\ \hline
%			Average OAP & 21.827 nm & 11.661 nm & 1.178 nm \\ \hline
%			DM & 11.2 nm & 3 nm & N/A \\ \hline
%			\end{tabular}
%		\end{center}
%	\caption{Caption}
%	\label{tab:OAPtable}
%\end{table}

\subsection{Software and Calibration}

CACTI uses the Compute And Control For Adaptive Optics (CACAO) real-time control software package \cite{guyon2018compute}. The calibration of the AO system by CACAO is a multi-step process. First, the hardware latency between the deformable mirror and wavefront sensor is measured to ensure proper synchronization. Then, the system response is measured by applying Hadamard modes to the DM and recording the corresponding WFS response. Using Hadamard modes to calibrate the AO system provides a higher SNR measurement of the interaction measurement than other modal basis sets such as Zernike polynomials\cite{kasper2004fast}. We calibrated the system with 0.01 $\mu m$ (1/63.3 $\lambda$) amplitude Hadamard modes to ensure that the system was calibrated within the linear range of the PWFS. CACAO saves the poke response matrix after each calibration. These files were analyzed to ensure the calibration was done correctly and at high SNR. To maximize the SNR of low-order modes the first few Zernike modes and low-order Fourier modes are used in combination with Hadamard modes in the AO system calibration. The Hadamard-encoded response is then decoded into a zonal response matrix, serving as a convenient intermediate step so that the system response can easily be projected to any modal basis. Fourier modes are adopted as a starting point for the modal wavefront control. We are unable to fully correct for all of the modes we applied during the calibration process, as aliasing, vibrations, misalignments, not well sensed modes affect the modes that can controlled by the AO system.

In these steps, the illumination pattern on the deformable mirror is determined by thresholding actuators according to the magnitude of their corresponding response in the wavefront sensor signals. Non-illuminated actuators are slaved to nearby illuminated actuators instead of being independently controlled. Similarly, a mask of the PWFS detector pixels is generated to define valid pixels from the PWFS pupils on the detector that will be used for wavefront sensing. 

CACAO is also used to generate phase screens to simulate turbulence. The streaming of turbulence is done independently of the AO loop processes and simulates a dynamic constantly evolving turbulent phase screen. The simulated wind speed is 17.4 mm/s across the 7.5mm beam diameter. Due to the limited low-order stroke of the BMC1K, the power spectrum of the turbulence generated by CACAO is filtered with a Gaussian unsharp mask to attenuate the low-order modes so that the full stroke of the DM is not used. At middle to high spatial frequencies, the power spectrum matches that of Kolmogorov turbulence. CACAO filters the low order modes to 89.7$\%$ of the total variance over our illuminated DM surface. Using this value, and following the methodology in Noll\cite{noll1976zernike} we can calculate $D/r_0$ or the turbulence screens generated by CACAO before attenuation. Table \ref{tab:Rtable} summarizes the unattenuated $D/r_0$ values for the turbulence strengths used in the experiments detailed in this paper. 

\begin{table}
	\begin{center}
		\begin{tabular}{ | l| l |l|}
			\hline
			\textbf{RMS Surface Error [microns]}&  \textbf{Unattenuated $D/r_0$} [at 633 nm] & \textbf{Greenwood frequency $f_G$}\\ \hline
			0.1 & 2.5 & 2.48 \\ \hline
			0.2 & 5.9 & 5.84\\ \hline
			0.3 & 9.5  & 9.41\\ \hline
			0.4 & 13.5 & 13.37\\ \hline
			0.5 &  17.6 & 17.44\\ \hline
			0.6  &  21.9 & 21.70\\ \hline
			\end{tabular}
		\end{center}
	\caption{RMS surface error in microns reported by CACAO for turbulence generation and the corresponding $D/r_0$ values before attenuation to not use the full stroke of the DM. The greenwood frequency for each turbulence strength is also detailed.}
	\label{tab:Rtable}
\end{table}

\subsection{AO Error Budget}

\begin{figure}
    \centering
    \includegraphics[width=0.7\textwidth]{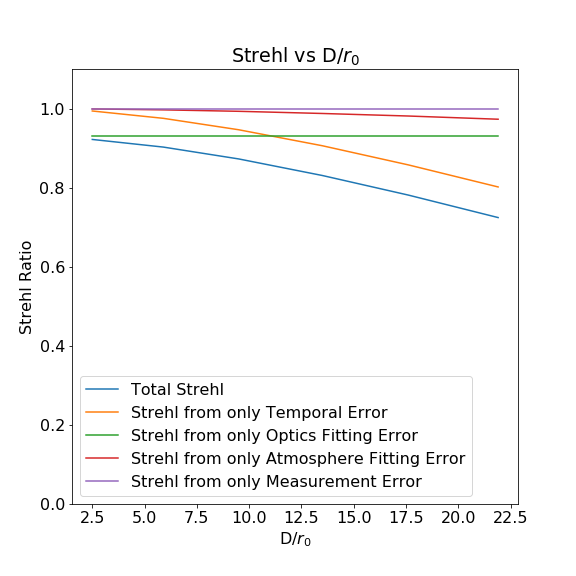}
    \caption{Estimated Strehl Ratio for each of the $D/r_0$ values simulated in CACTI. The total Strehl value is the highest estimated closed-loop Strehl value achievable on CACTI when considering all of the terms in the error budget. The other lines are plots of the expected Strehl values when only considering a single term from the error budget. }
    \label{fig:errorbudget}
\end{figure}

Here we present the terms of the AO error budget for the CACTI adaptive optics simulator based on values determined by our system calibration and characterization. The deformable mirror in CACTI is used to both apply and correct atmospheric turbulence. We are unable to fully correct for all the modes we apply, and therefore have a fitting error that is dependent on the value of the turbulence parameter $r_0$. We calculated for each value of $r_0$ the fitting error using the formula for fitting error from Hudgin\cite{hudgin1977wave}. An additional static fitting error exists in CACTI from the surface error of our optics. We correct up to 6 cycles per aperture of surface error using the DM. From our PSD analysis in Section \ref{characterization} we determined that we have an estimated Strehl Ratio of 0.93 based on the optical surface errors in CACTI. In CACTI we have a temporal error that is dependent on the Greenwood frequency of turbulence and our lag time. The time lag in CACTI is 1 frame, which corresponds to 0.0025 seconds. Using the equation for temporal error from Hardy\cite{hardy1998adaptive}, we calculated the temporal error terms for each strength of turbulence used in CACTI. In addition, we expect a measurement error term that we estimate using $\beta_p$, the sensitivity of the wavefront sensor to photon noise\cite{guyon2005}. The equation used for measurement noise is,
\begin{equation}
    \sigma^2=(\beta_p^2/N_{photon})*k
\end{equation}

\noindent where $k$ is the spatial frequency content measured by the wavefront sensor. The variable $N_{photon}$ is the number of photons available for wavefront sensing, which we estimated using a 1000 frame average of our wavefront sensor signal to a flat wavefront and the average gain value for our detector pixels. Figure \ref{fig:errorbudget} displays the estimated Strehl value for each value of $D/r_0$ that we simulate for the CACTI system. The blue line is the estimated highest AO closed-loop Strehl achievable on CACTI when considering all of the error terms. We have also included the values of Strehl when only considering individual error terms.

\subsection{Pyramid Wavefront Sensor Testbed}

\begin{figure}
    \centering
    \includegraphics[width=0.8\textwidth]{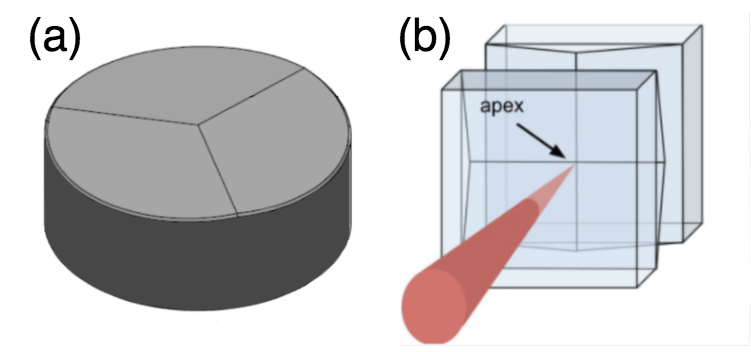}
    \caption{Drawings of the pyramid optics in CACTI. A. The 3PWFS pyramid is a single prism made from fused silica glass. The 4PWFS pyramid is two crossed roof prisms. This pyramid is a copy of the pyramid used by SCExAO. This drawing is Figure 2. from Jovanovic et al.\cite{jovanovic2016scexao}. }
    \label{fig:pyramidOptics}
\end{figure}

The design of the PWFS testbed is detailed in Figure \ref{fig:CACTI}. Light from the AO simulator is collimated by a 500 mm focal length lens (L1) which relays the exit pupil of the AO simulator to the PWFS testbed. The exit pupil of the AO simulator becomes the entrance pupil to the PWFS testbed, which is then resized by a pupil relay consisting of an achromatic doublet (L2) and a custom air-spaced achromatic triplet lens (L3). The pupil is imaged on the modulation mirror, which is a $\lambda/20$ flat mirror mounted on a piezo-actuator tip/tilt platform. A 3PWFS designed by HartSCI L.L.C. has been integrated into the instrument for the performance test. The three-sided pyramid optic is a single prism made from fused silica glass and has an excellent tip smaller than 5 microns. Figure \ref{fig:pyramidOptics}.A shows the 3D model of the manufactured prism. The 4PWFS in CACTI uses two crossed roof prisms for its pyramid optic shown in Figure \ref{fig:pyramidOptics}.B. This is the same type of pyramid used in the SCExAO\cite{jovanovic2015subaru}.  A 50/50 beamsplitter after the modulation mirror sends the same F/53 focused beam to the pyramid tips of the 3PWFS and 4PWFS to mitigate non-common path errors, as all the optics up to that point are common to both wavefront sensors.

%The other optics in the CACTI system are off the shelf lenses quoted at a surface quality of $\lambda/4$ PV which translates to an RMS surface of 39.56 nm following the convention set by Schwertz\cite{schwertz2010useful} at a wavelength of 633 nm. 

The pupil of each PWFS is imaged on the detectors by using two camera lenses C1 and C2 for the 3PWFS, C3 and C4 for the 4PWFS that form a zoom lens system to ensure that the sizes of pupils from both PWFS are 30 pixels in diameter. This diameter slightly oversamples our deformable mirror which has 28 actuators in X and 26 actuators in Y illuminated.  Using this setup we were able to align the pupils to be 30.5 pixels across in the 4PWFS arm, and 29.5 pixels across in the 3PWFS arm. The 3PWFS uses an sCMOS, and the 4PWFS uses a CMOS detector. The experiments performed in this paper were under bright light conditions, so having different cameras with different noise characteristics should not affect the PWFS performance as we are not read noise limited. The measurements were taken at high SNR above the read noise limit and well within the linear range of the detectors. The average count per pixel for a 3PWFS pupil was 1188 counts out of a well depth of 4096 counts. This count was determined by applying a flat on the DM and calculating the average pixel value for only the pixels within the pupils over 1000 frames. 

The PWFS signals on CACTI can be processed in two ways: the Raw Intensity (RI) and Slopes Maps (SM) methods. In both methods, the detector signal from the PWFS is dark subtracted, and a threshold is applied to mask out any pixels outside of the PWFS pupils. A reference frame that is the recorded pupil intensity values is then subtracted off of the signal. For the SM method a slopes reference is subtracted off instead. The intensity values in each of the pixels are normalized by the sum of the intensities in all of the pupils. The remaining signal is now only the intensity pattern due to the phase error. The Raw Intensity method uses this intensity signal as is, and extracts the remaining signal into a 1D column vector. The Slopes Maps calculation recombines the PWFS pupils into an estimate of the X and Y slopes of the wavefront slope. The SM equation for the 4PWFS is given in Equation \ref{4PWFSslopes}, and the Equation for the 3PWFS is given in Equation \ref{3PWFSslopes}. In these equations $S_x, S_y$ are the local wavefront slopes, and $I_1, ... ,I_4$ are the intensity values of the pixel corresponding to the same location in each pupil.

\begin{eqnarray}
    S_x=\frac{I_1+I_2-I_3-I_4}{I_1+I_2+I_3+I_4}     \label{4PWFSslopes} \\
    S_y=\frac{I_1-I_2-I_3+I_4}{I_1+I_2+I_3+I_4} \nonumber
\end{eqnarray}

\begin{eqnarray}
    S_x=\frac{\frac{\sqrt{3}}{2}I_2-\frac{\sqrt{3}}{2}I_3}{I_1+I_2+I_3} \label{3PWFSslopes} \\
    S_y=\frac{I_1-\frac{1}{2}I_2-\frac{1}{2}I_3}{I_1+I_2+I_3} \nonumber
\end{eqnarray}
% The PWFS signals on CACTI were processed using the the Raw Intensity (RI) method. The detector signal from the PWFS is dark subtracted, and a threshold is applied to mask out any pixels outside of the PWFS pupils. In the RI method the remaining signal is used as-is and are unraveled into a vector of intensity values that is used for calibration and wavefront reconstruction.

A concern for the 3PWFS is that the wavefront is sampled with three points instead of four, and thus potentially has a larger null space than the 4PWFS. This would mean fewer modes are sensed, and the accuracy of the wavefront sensing is reduced. On the CACTI testbed, we found that the difference in sampling did not impact system performance, or affect the number of modes the AO was able to close on. A summary of the number of basis set modes that were used to close the loop for each PWFS and signal processing method can be found in Table \ref{tab:Modestable}. The number of modes used is determined by our control software during the AO system calibration. Although the number of modes varies between PWFS, modulation radius, and signal handling method, these differences are slight. Following the analysis presented in Noll\cite{noll1976zernike}, we can calculate the expected mean square residual wavefront error after correction by a number of wavefront modes. Examining the residual error from the least (493) and highest (516) modes corrected in our system we find that the difference in residual error between the two modes for the same $D/r_0$ is 0.00005 [$rad^2$] which is negligible to the impact of Strehl performance. 

\begin{table}
	\begin{center}
		\begin{tabular}{ | l|l|l | l| }
			\hline
			\textbf{PWFS}& \textbf{Signal Processing} &\textbf{Modulation radius} &\textbf{$\#$ of Modes}\\ \hline
             3PWFS & Raw Intensity & 3.25 $\lambda/D$ & 515\\ \hline
             3PWFS & Raw Intensity & 1.6 $\lambda/D$ & 507 \\ \hline
             3PWFS & Slopes Maps &  3.25 $\lambda/D$ &493 \\ \hline
             3PWFS & Slopes Maps &  1.6 $\lambda/D$ & 496\\ \hline
             4PWFS & Raw Intensity & 3.25 $\lambda/D$ & 516\\ \hline
             4PWFS & Raw Intensity & 1.6 $\lambda/D$ & 513\\ \hline
             4PWFS & Slopes Maps &  3.25 $\lambda/D$ & 497 \\ \hline
             4PWFS & Slopes Maps &  1.6 $\lambda/D$ & 502\\ \hline
			\end{tabular}
		\end{center}
	\caption{Number of basis set modes used for the AO closed loop for each PWFS and signal processing method.}
	\label{tab:Modestable}
\end{table}

\subsection{Current Status}

\begin{figure}
    \centering
    \includegraphics[width=0.8\textwidth]{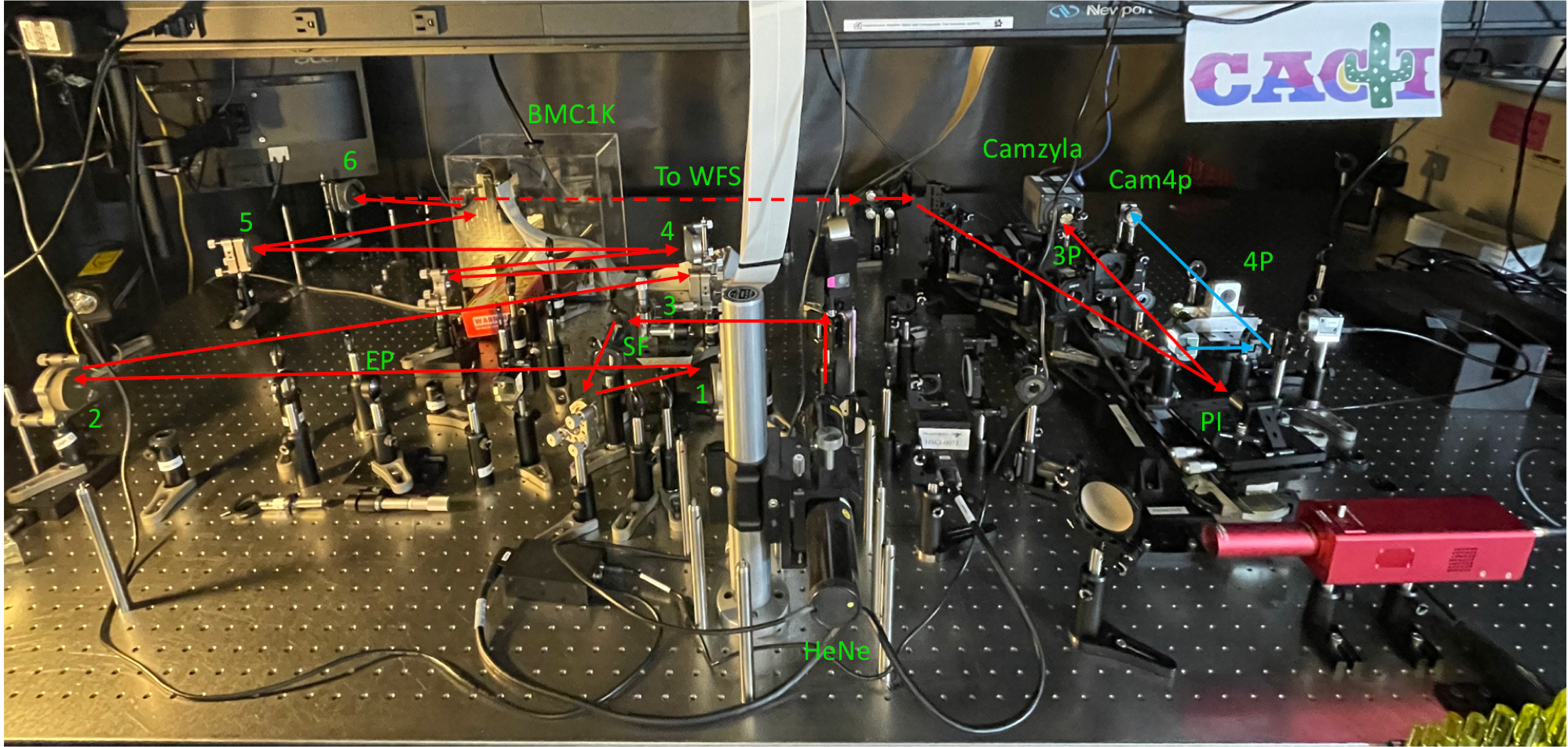}
    \caption{Image of the as-built CACTI testbed. Light starts from the HeNe laser and propagates through the AO simulator on the left half of the optical table. After the BMC1K and the final OAP, the light is relayed to the PWFS testbed on the right half of the optical table.}
    \label{fig:cactiTestbed}
\end{figure}

The alignment of the adaptive optics simulator and PWFS tested in CACTI was completed in May 2020. Figure \ref{fig:cactiTestbed} is a picture of the as-built system. For a detailed schematic of the CACTI optical layout see Figure \ref{fig:CACTI}. The measured pupil diameter for the 4PWFS is 30.5 pixels and 29.5 pixels for the 3PWFS. Both of the PWFS detectors are oversampled with respect to the 27.2 actuators across the deformable mirror pupil we determined in calibration. In addition we find in our calibration process that the number of modes corrected by each PWFS is almost the same. We conclude therefore that the pixel difference in sampling does not noticeably change the relative performance of the wavefront sensors.

%Light from the HeNe laser is relayed by the six OAP mirrors and is then coupled to the PWFS testbed using a beam triangle formed by two flat mirrors. Figure \ref{fig:PWFStestbed} is a close-up diagram of the PWFS testbed. The first two lenses (L1 and L2) relay the entrance pupil of the PWFS testbed onto the PI modulation mirror (PI). A beamsplitter after the modulation mirror picks off the light to the four-sided pyramid (4P) and the through-beam is sent to the three-sided pyramid (3P). Both pyramids have two camera lenses (C1 and C2) that form a zoom lens to match the diameters of the pupils of the two PWFS. 

\begin{figure}
    \centering
    \includegraphics[width=0.8\textwidth]{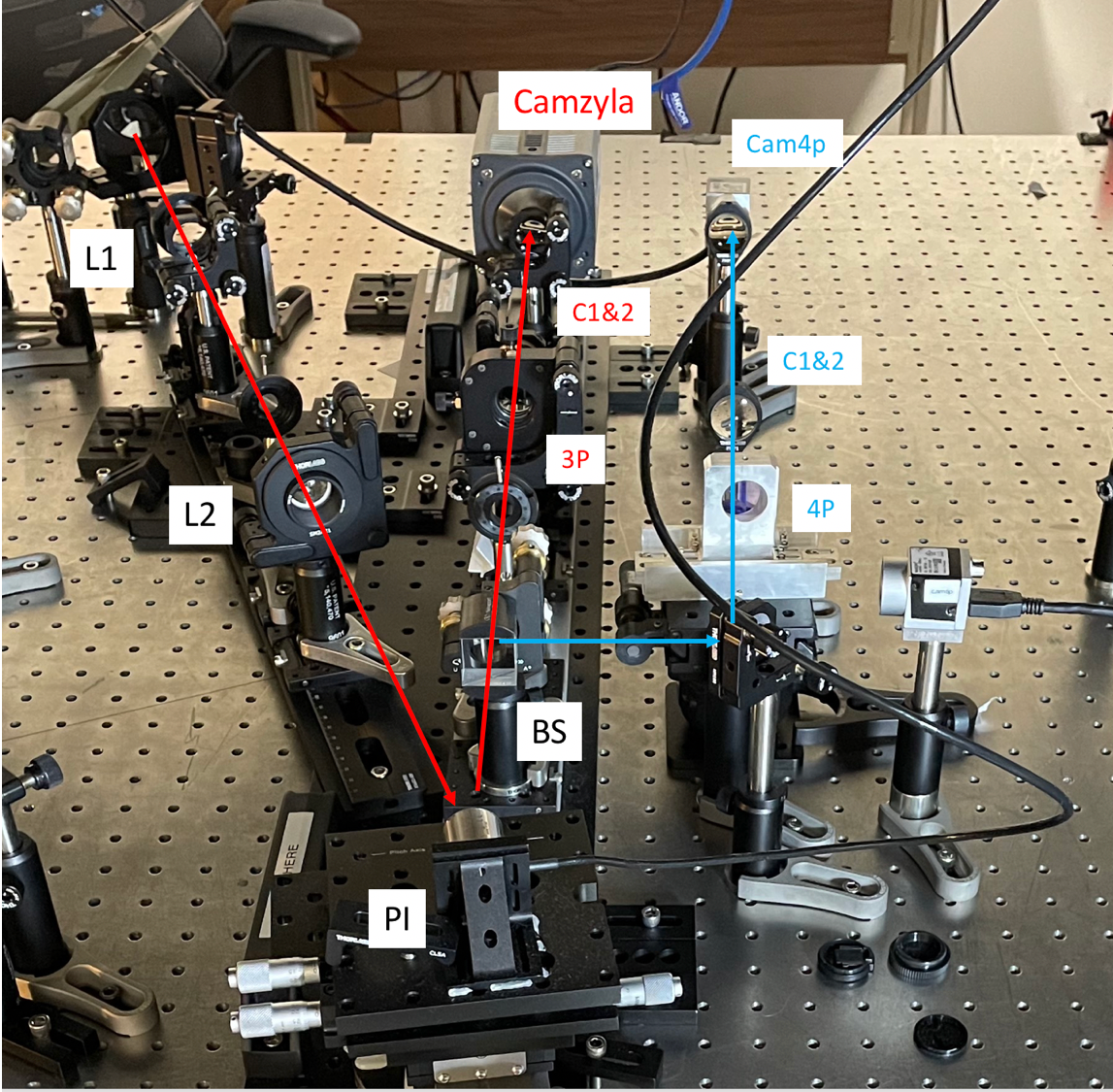}
    \caption{Close up image of the PWFS testbed. Light enters the system before L1 on the left side. The light is relayed onto the modulation mirror (PI). The through beam propagates through to the 3PWFS which consists of the pyramid optic (3P), camera lenses (C1$\&$2) and the sCMOS camera (Camzyla). Light is by a beamsplitter (BS) to the 4PWFS arm which consists of the pyramid (4P), camera lenses (C3$\&$4) and the CMOS camera (Cam4p).  }
    \label{fig:PWFStestbed}
\end{figure}

\begin{figure}
    \centering
    \includegraphics[width=0.8\textwidth]{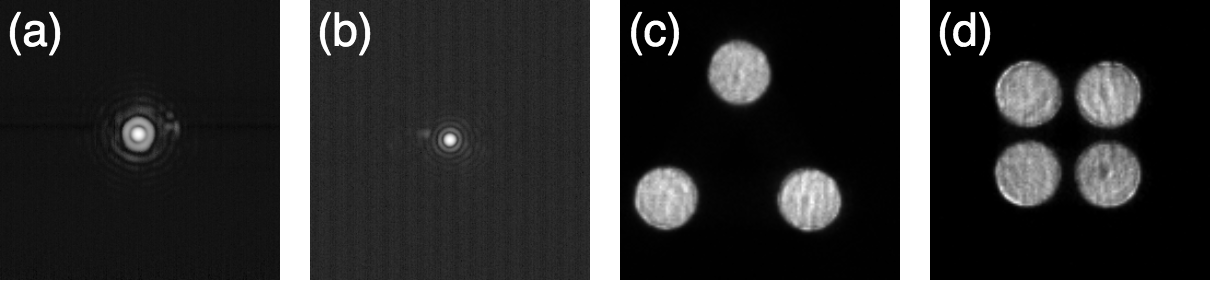}
    \caption{Signals from CACTI in response to the optimized flat wavefront. A. The PSF on the science camera. B. The optimized PSF on the focal plane of the PWFS tips. C. The 3PWFS pupils on Camzyla at 5 $\lambda/D$ modulation. D. The 4PWFS pupils on Cam4p at 5 $\lambda/D$ modulation.}
    \label{fig:flatCACTI}
\end{figure}

 The system responses to the wavefront flattened by the deformable mirror are given by Figure \ref{fig:flatCACTI}. Figure \ref{fig:flatCACTI}.A is the PSF on our science camera on a logarithmic color scale. Figure \ref{fig:flatCACTI}.B similarly is the PSF on the pyramid tip also in log scale. Figure  \ref{fig:flatCACTI}.C and Figure \ref{fig:flatCACTI}.D are the pyramid pupils from a flat wavefront with 5$\lambda/D$ modulation for the 3PWFS and 4PWFS respectively. There is a ghost in the CACTI PWFS located at the \nth{4} Airy ring, caused by one of the lenses in the system.  Both PWFS see the same PSF and are effected in the same way. The consequence of the ghost is to cause a sinusoid pattern in the intensity signal of the PWFS signals that is removed in calibration.  We operate the PWFS in bright light conditions and are not affected by the reduction in dynamic range caused by the static sinusoid pattern. Using a model of the CACTI PSF we calculated the Strehl Ratio for the PSFs in Figure \ref{fig:flatCACTI}.A and \ref{fig:flatCACTI}.B. The Strehl Ratio on our optimized pyramid tip focal plane (Figure \ref{fig:flatCACTI}.B) was found to 0.90, which is in agreement with our PSD analysis which predicted an ideal Strehl Ratio of 0.93 for the CACTI system with only static fitting error from the optical surfaces present. Our science PSF (Figure \ref{fig:flatCACTI}.A) which does not benefit from PSF sharpening was found to have a Strehl Ratio of 0.7. 

The PSF on the pyramid tips were sharpened by the DM using the eye-doctor algorithm previously described. The three-sided pyramid and the four-sided pyramid see the same PSF, optimizing for one optimizes the PSF for both. To sharpen this PSF we inserted a fold flat mirror right before the pyramid optics and sent the PWFS focal plane to a camera. A feedback loop is formed between the DM and the PSF detected on this camera. This optimization was done to insure that each PWFS saw exactly the same PSF, and that the common path errors up to the pyramid tip are compensated for. The high Strehl PSF on the pyramid tips ensures that both PWFS are operating in the linear range. The consequence of this is that the PSF at the science plane is not optimized. The goal of this experiment is not to determine the maximum Strehl value a pyramid wavefront sensor can produce. Rather we seek to demonstrate the operation of a 3PWFS and prove that it can provide wavefront correction of turbulence good enough to return the PSF close to its original state.

We have successfully closed the AO loop on CACTI with both the 4PWFS and 3PWFS, using both the RI and SM signal processing methods. This marks the first time that the AO loop has been closed on a 3PWFS employing a glass pyramid. Previous work by Schatz et al.\cite{schatz2021three} closed the very first AO loop on a 3PWFS on the LOOPS testbed at the Laboratoire d'Astrophysique de Marseille, where the three-sided pyramid was created by a phase screen applied to a spatial light modulator.

\begin{figure}
    \centering
    \includegraphics[width=0.8\textwidth]{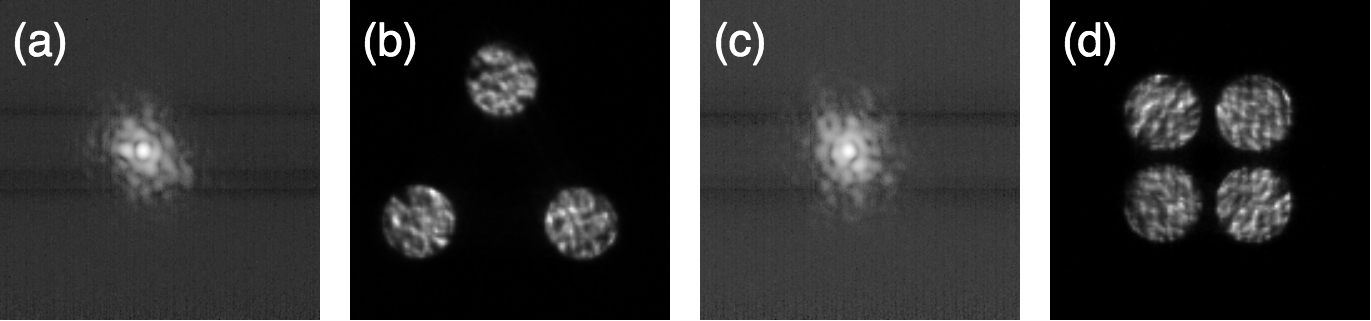}
    \caption{A. Time-averaged science camera image of the 3PWFS closed-loop PSF in log scale. B. Turbulence streaming across the 3PWFS pupils. C. Time-averaged science camera image of the 3PWFS closed-loop PSF in log scale. D. Turbulence streaming across the 4PWFS pupils.}
    \label{fig:turbCACTI}
\end{figure}

The deformable mirror on CACTI was used to both generate the turbulence screen and apply the correction. This was done using the CACAO software, which creates multiple channels of commands for the deformable mirror. In one channel, we can stream the turbulence phase screens, and in another channel, we have the commands computed by real-time control software using the PWFS signals. The actual command applied to the DM is a summation of commands from all of the channels. Figure \ref{fig:turbCACTI} shows the closed-loop PSFs and pyramid pupils from a turbulence strength of 0.3 $\mu$m RMS error on CACTI. Figure \ref{fig:turbCACTI}.A and \ref{fig:turbCACTI}.B are the 3PWFS pupils and the closed-loop PSF in log scale on the science camera. Figure \ref{fig:turbCACTI}.C and \ref{fig:turbCACTI}.D are the 4PWFS pupils and the closed-loop PSF in log scale on the science camera.

% \jrmcom{GREAT STUFF!!!}

% \begin{table}
% 	\begin{center}
% 		\begin{tabular}{ | l|l|l | }
% 			\hline
% 			\textbf{Camera}& \textbf{Model} &\textbf{Read Noise}\\ \hline
%              Camsci & Basler ACE acA640-750um CMOS  & 12.25 $e^-$ read noise\\ \hline
%              Camzyla & Zyla 4.2+ sCMOS detector & 7.98 $e^-$ read noise \\ \hline
%              Cam4p & Basler ace acA720-520um CMOS & 0.11 $e^-$ read noise \\ \hline
%              Camtip & Basler ACE acA640-750um CMOS (DOUBLE CHECK)  & 9.30 $e^-$ read noise\\ \hline
				
% 			\end{tabular}
% 		\end{center}
% 	\caption{CACTI components}
% 	\label{tab:CACTItable}
% \end{table}

\section{Experimental Details}

The CACTI testbed was used to compare the performance of a 3PWFS and 4PWFS in varying strengths of turbulence. Previous work by Schatz et al.\cite{schatz2021three} found in simulation that the performance of the two wavefront sensors is comparable. These simulations however were run for only one seeing condition. The goal of this experiment is to determine the relative performance of the 3PWFS to the 4PWFS in varying strengths of turbulence using both the Raw Intensity and Slope Maps signal processing methods. The performance was determined by measuring the relative Strehl ratio of the AO corrected PSF and the aberration-free PSF. For the CACTI testbed, our aberration-free PSF was the flat wavefront PSF on the science camera created by the deformable mirror shown in Figure \ref{fig:flatCACTI}.A., which we will refer to as the reference PSF. The reference PSF was used in the calibration of each PWFS. The relative Strehl ratio is calculated by taking the ratio of the peak intensity of the partially corrected PSF, $P_{data}$,  to the peak intensity of the reference PSF, $P_0$. The peaks are normalized by the flux in the image. 

\begin{equation}
    S=\frac{P_{data}/Flux_{data}}{P_{0}/Flux_{0}}
    \label{Strehl}
\end{equation}

The experiment performed on CACTI was to measure the relative Strehl Ratio as a function of turbulence strength and modulation radius for both the 3PWFS and the 4PWFS.  This experiment was performed using both the Slopes Maps and Raw Intensity signal processing methods. The modulation radii used were: 1.6 $\lambda/D$, and 3.25 $\lambda/D$. At each modulation radius the following experiment was performed at a loop speed of 400-Hz:

\begin{itemize}
    \item The best flat commands were applied on the DM and a reference PSF was calculated from the average of 1000 frames.
    \item Dark frames were taken for each of the cameras and subtracted from each PWFS frame as a signal processing step in the closed-loop. These dark frames were created by turning off all light sources and taking average of 1000 frames from the detector.
    \item Reconstruction matrices were calculated by taking system calibrations for each modulation radius.
    \item Simulated turbulence screens were generated at RMS wavefront error values of 0.1 $\mu$m, 0.2 $\mu$m, 0.3 $\mu$m, 0.4 $\mu$m, 0.5 $\mu$m, and 0.6 $\mu$m. The simulated wind speed is 17.4 mm/s across the 7.5mm beam diameter.
    \item At each phase screen 50 closed-loop PSF images were recorded. Each of these 50 images was created by taking the average from 300 frames of data. 
    \item The average Strehl value from the closed-loop PSFs was calculated to create an average Strehl value for each turbulence strength.
    \item Plots were then created of Strehl value as a function of both turbulence strength and modulation radius. 
    
\end{itemize}

The loop gain for the experiments was set to 0.8. This is higher than normal AO operations on-sky. On CACTI we generate our turbulence with the DM. In each step of the AO loop a new phase screen is applied. In addition we operate in bright light conditions. Both of these factors enable us to close the loop with a high loop gain. CACAO allows for gains to be set for blocks of spatial frequencies. For example, modal block 00 controls the gain of Tip/Tilt, and block 01 controls Focus. It is necessary to optimize modal gains rather than use a single value of loop gain due to the sensitivity and optical gain of the PWFS. Schatz et al. 2021\cite{schatz2021three} found that the sensitivity of an unmodulated PWFS operating in the linear, infinite pupil regime is independent of spatial frequency. In a real PWFS system, the modulation of the PWFS and the finite size and sampling of the pupils modifies the sensitivity of the PWFS. Correia et al.\cite{correia2020performance} plot the sensitivity function of the PWFS which is dependent on the spatial frequency and radius of modulation. The sensitivity of the wavefront sensor is also dependent on the atmospheric seeing. In stronger turbulence, the spot on the pyramid tip is degraded by uncorrected residual turbulence. The so-called optical gain of the pyramid is a spatial frequency-dependent term that captures the change between the in-lab calibrated PWFS sensitivity and the on-sky sensitivity. Chambouleyron et al. (2020) \cite{chambouleyron2020pyramid} shows the changes in the optical gain of the PWFS for different modulation and radii and turbulence strength. 

\begin{figure}
    \centering
    \includegraphics[width=0.7\textwidth]{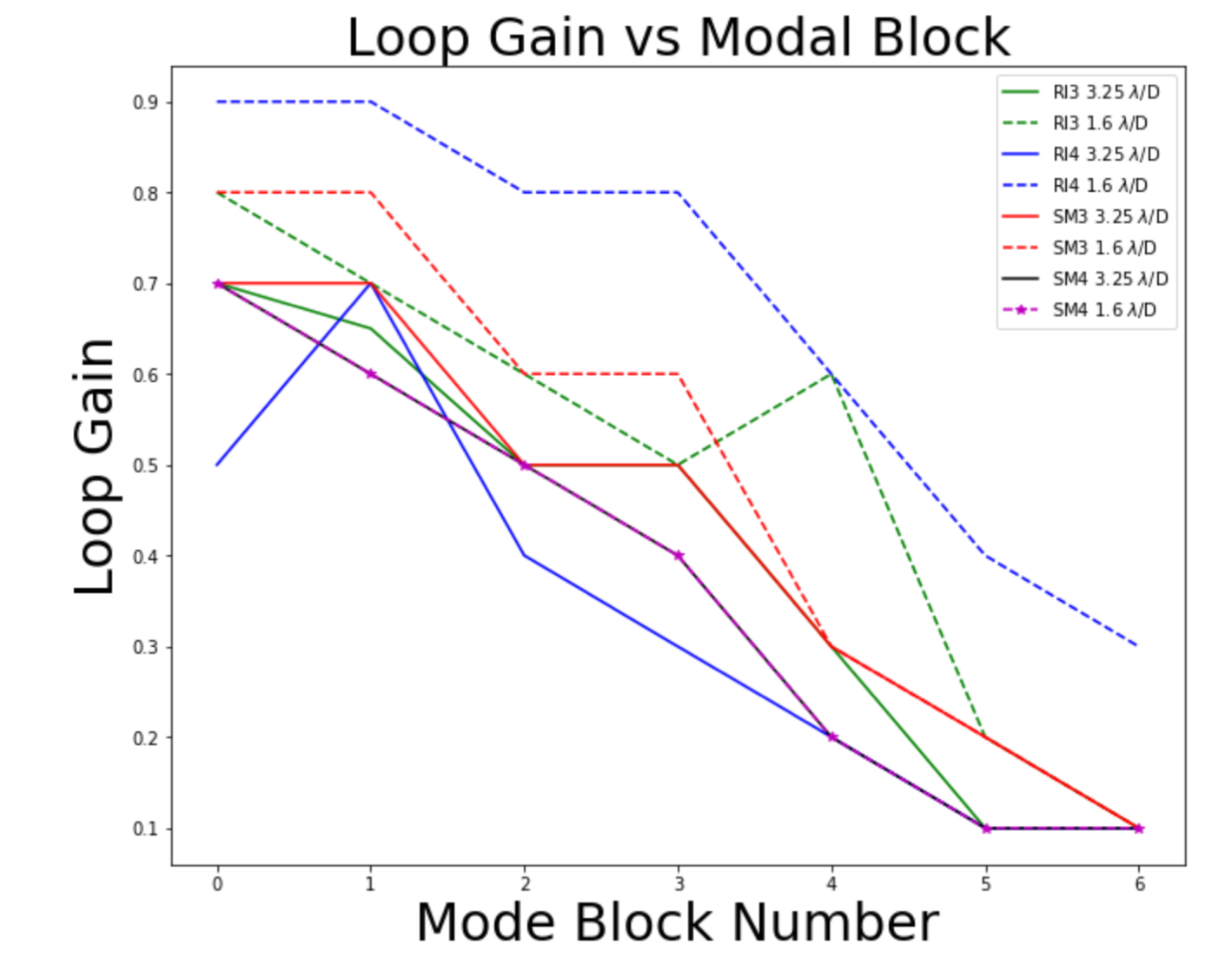}
    \caption{Loop gain value optimized for the PWFS configuration plotted against the corresponding modal block number in CACAO. The modal gains for SM3 3.25 $\lambda/D$ and SM4 1.6 $\lambda/D$ were the same. To differentiate we have over plotted the SM4 1.6 $\lambda/D$ line in magenta dashes and stars. The gain was optimized at low turbulence, and the same values were used for all turbulence strengths.}
    \label{fig:gains}
\end{figure}

Actively optimizing the modal gain of the PWFS to compensate for the PWFS sensitivity and changing optical gain is a necessary step for on-sky PWFS optimization. Chambouleyron et al. (2021)\cite{chambouleyron2021focal} provides a methodology of actively estimating the optical gain of the PWFS on-sky by monitoring the PSF on the pyramid tip. For this experiment, the modal gains were optimized for each PWFS configuration by performing a crude search. At high levels of turbulence, the modal gains were tuned to maximize the AO system correction by eye. PSF frames were recorded and a Strehl ratio was calculated for those gain values. The gains were then adjusted and the same procedure was performed until the values of the modal gains converged to give a good correction. These optimized gain values were then used for all levels of turbulence. This method, although not the optimum, has been used used for on-sky AO systems. Our goal is not to measure the highest Strehl achievable for each PWFS but to prove that the 3PWFS can reconstruct the wavefront as well as a 4PWFS under similar conditions. Figure \ref{fig:gains} plots the modal loop gain value against the modal frequency block number. The modal gains for SM3 3.25 $\lambda/D$ and SM4 1.6 $\lambda/D$ were the same. To differentiate we have over plotted the SM4 1.6 $\lambda/D$ line in magenta dashes and stars. In general lower modulation allowed for higher loop gains. All wavefront sensors configurations show a downward slope to the loop gain. The differences in the values between PWFS configurations is most likely a reflection of our imprecise gain tuning. Accurate optimization of the modal loop gains is a subject for future work.

%An adaptive optics system works on the concept of conjugate imaging. An object plane corresponding to a layer of atmospheric turbulence is imaged through the telescope AO system onto a deformable mirror where the correction is applied. To achieve this the bulk of an AO system on sky or in the lab consists of pupil relays.

% \begin{figure}
%     \centering
%     \includegraphics[width=1\textwidth]{psfs.png}
%     \caption{PSFs on the science camera with A) Flat Wavefront, B) Turbulence, C) AO Corrected Turbulence.}
%     \label{fig:PSF}
% \end{figure}

% \begin{figure}
%     \centering
%     \includegraphics[width=0.5\textwidth]{3poptic.png}
%     \caption{The 3D model of the refractive 3 side pyramid optic.}
%     \label{fig:3poptic}
% \end{figure}

\section{Results}

\begin{figure}
    \centering
    \includegraphics[width=0.8\textwidth]{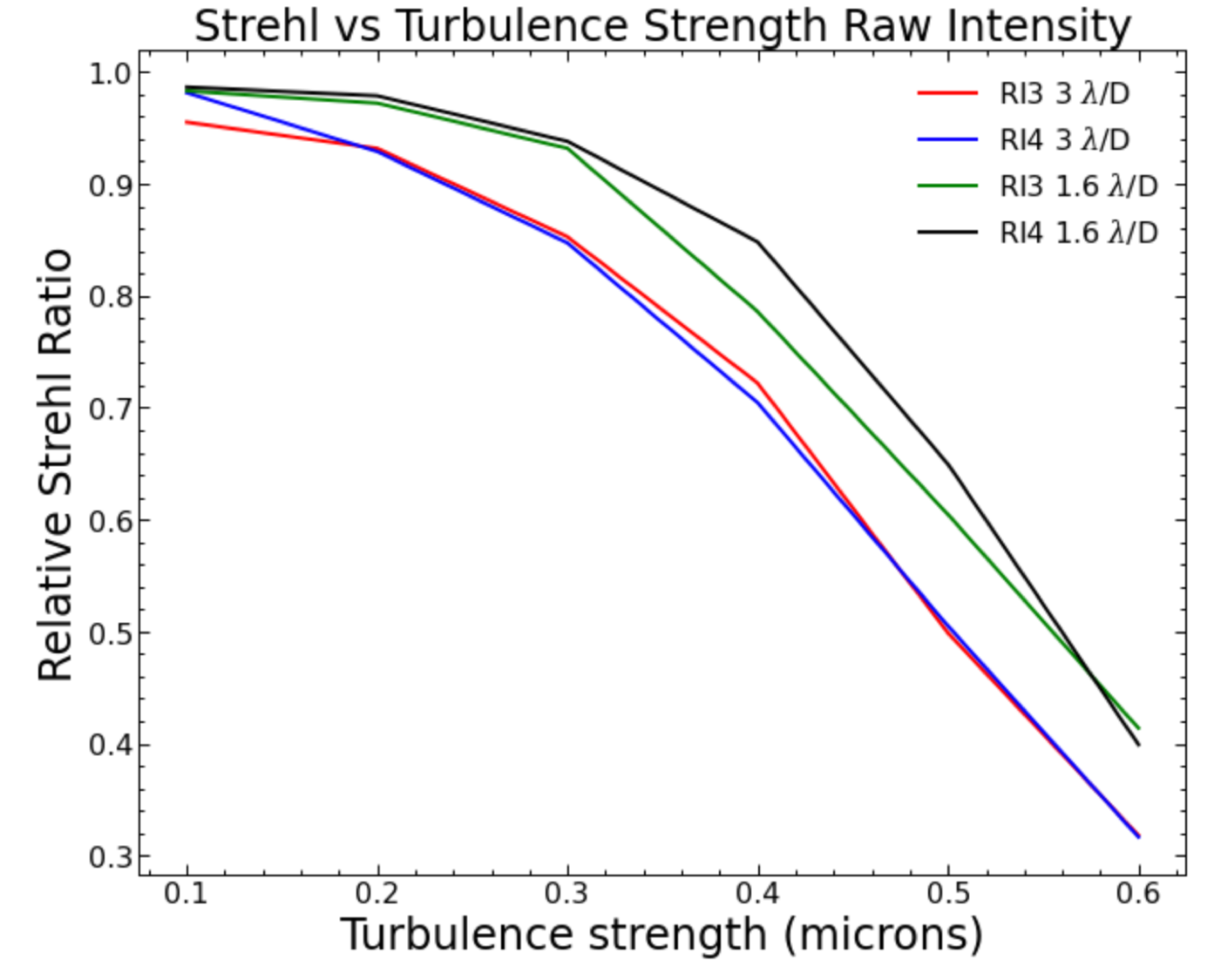}
    \caption{Relative Strehl ratio versus turbulence strength for the 3PWFS and 4PWFS using the Raw Intensity signal processing method. The performance of each wavefront sensor at both 3.25 $\lambda/D$ and 1.6 $\lambda/D$ are comparable. A slight increase in performance is seen when moving to lower modulation. }
    \label{fig:RI}
\end{figure}

We found that the 3PWFS and 4PWFS were comparable at all levels of turbulence and modulation. Figure \ref{fig:RI} plots the resulting relative Strehl ratio curves for varying turbulence strengths for both the 3PWFS and 4PWFS using the Raw Intensity signal processing method. There is a slight increase in performance when the system has lower modulation. The sensitivity of the PWFS increases with lower modulation at the cost of dynamic range. In CACTI we filter out lower-order modes which have the most power in the atmospheric turbulence PSD. We therefore benefit from the increase in sensitivity instead of losing performance due to limited dynamic range. This has implications that a highly-sensitive ExAO system can be built by introducing a low-order AO loop upstream of the PWFS. However, the modal gain optimization was crude, and this effect could be a reflection of imperfect gain selection for the 3.25 $\lambda/D$ case. 

\begin{figure}
    \centering
    \includegraphics[width=0.8\textwidth]{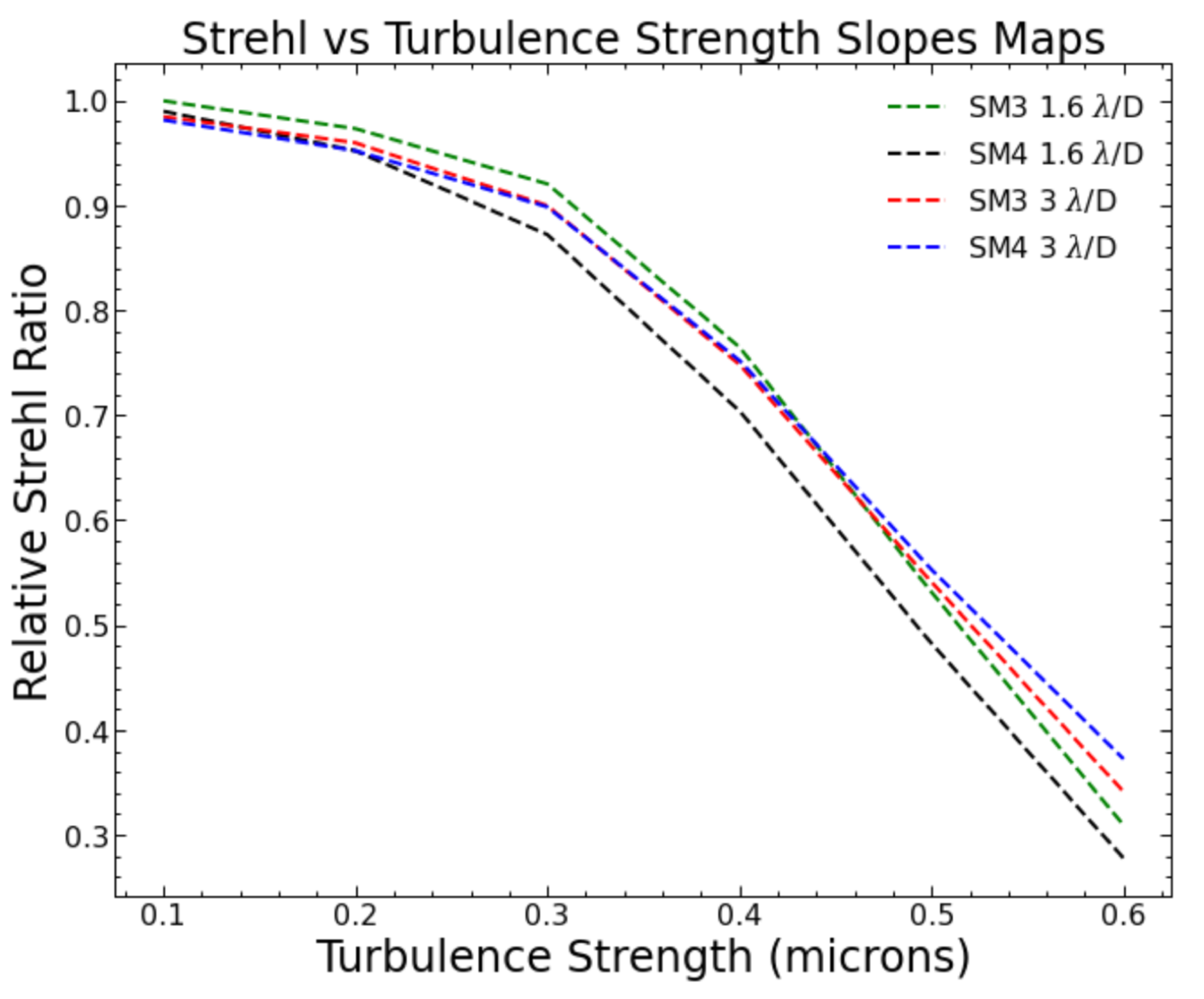}
    \caption{Relative Strehl ratio versus turbulence strength for the 3PWFS and 4PWFS using the Slopes Maps signal processing method. The performance of each wavefront sensor across all modulations radii and turbulence strengths are comparable.}
    \label{fig:SM}
\end{figure}

The performance of the PWFS using the Slopes Maps calculation was more uniform. The relative Strehl ratio calculated for the 3PWFS and 4PWFS at 3.25 $\lambda/D$ and 1.6 $\lambda/D$ are comparable. Figure \ref{fig:SM} plots the relative Strehl ratio as a function of turbulence strength for the PWFS using the Slopes Maps calculation.

\begin{figure}
    \centering
    \includegraphics[width=0.8\textwidth]{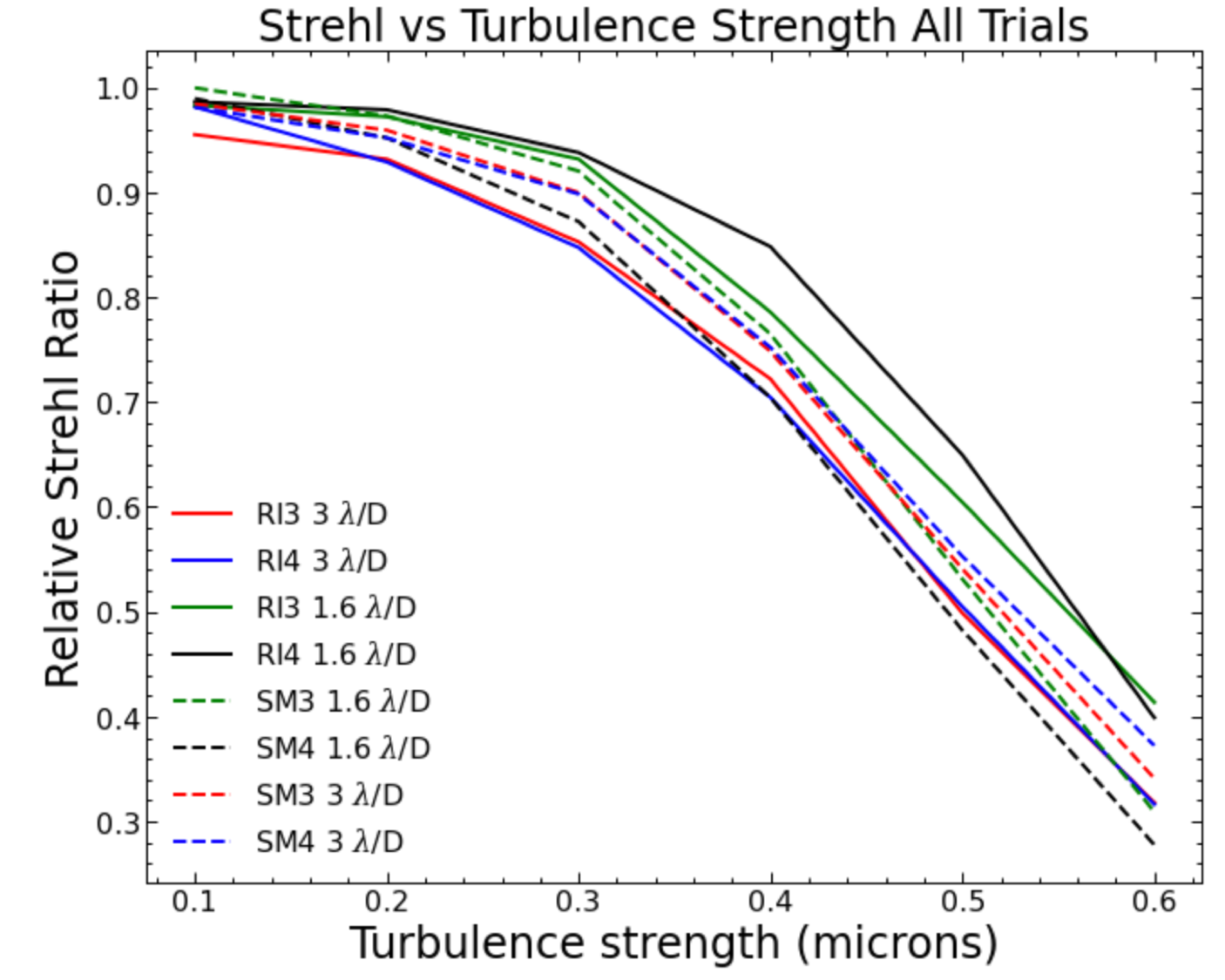}
    \caption{Summary of all results. Best performance was achieved using the 4PWFS at 1.6 $\lambda/D$ modulation using RI. However, this trial had the highest modal gains set, suggesting that this performance difference is not real, and that the modal gain for the other trials was set too low.}
    \label{fig:All}
\end{figure}

Figure \ref{fig:All} plots all of the results from both the RI and SM trials onto a single plot. This plot shows that the best performance was obtained by the 4PWFS using RI at 1.6 $\lambda /D$ modulation. Referring back to the plot in Figure \ref{fig:gains}, we can see that this trial also used the highest modal gains. This suggests that this increase in performance for the 4PWFS is not real, and that the modal gains for the other trials were not properly optimized and that increasing the modal gains could have increased performance.

\section{Discussion}

In the current configuration, a 3PWFS and 4PWFS were integrated into CACTI for a performance test. Both PWFS were designed with refractive pyramid optics and had similar sampling across the pyramid pupils. An effort was put into minimizing the differences between each PWFS. Non-common path error was minimized by ensuring each pyramid optic had the same PSF on the tip. Two different signal processing methods were employed on CACTI to process the PWFS signals.

The performance of each wavefront sensor on the CACTI testbed was determined to be similar for each modulation radius and turbulence strength. The difference in performance for the Raw Intensity method for the 1.6 $\lambda/D$ and 3.25 $\lambda/D$ modulation cases was on average 0.069 Strehl for the 3PWFS, and 0.086 Strehl for the 4PWFS. It has been shown \cite{guyon2005}$^,$ \cite{verinaud2004nature} that the sensitivity of the PWFS is increased by decreasing the radius of modulation. We cannot definitively conclude that the increase in performance of the 1.6 $\lambda/D$ modulation case was due to the increase in PWFS sensitivity due to systematic errors such as misalignment, imperfect calibrations, or imperfect selection of modal gains. The performance of each wavefront sensor was maximized when the modal gains were tuned. Modal gain tuning is therefore a necessary step to optimize the correction of an ExAO system. Nothing in our experiments indicated that with proper gain tuning the performances of the 3PWFS and 4PWFS should be different.

Comparing the results in Figure \ref{fig:All} to our AO performance prediction in Figure \ref{fig:errorbudget}, we that at high levels of turbulence the CATCI AO system greatly under-performs compared to our predictions. The cause of this is saturation of the deformable mirror inter-actuator stroke. The inter-actuator stroke of our DM is roughly $\pm$ 0.4 microns. 

% The power of low order modes of the turbulence screens used by CACTI are filtered so that the full stroke of the DM is not used. \jrmcom{This isn't quite right.  The power at low spatial frequencies is lower than it would be in unfiltered Kolmogorov turb, but it is still higher than at higher spatial frequencies.  That is, the filter does not set it to 0.} Most of the power in the spatial frequencies of the turbulence screen power spectrum are mid to high spatial frequencies. \jrmcom{So this statement might be too big:}Our results suggest that modulating decreases sensitivity across all spatial frequencies, and that the AO system should be run at as small of a modulation radius as possible.

The demonstration of the performance of the 3PWFS is a critical step in the development of the wavefront sensor. An attempt was made to compare the performance of the 3PWFS and 4PWFS in low light conditions on the CACTI testbed. This initial experiment lacked the precision to produce quantifiable results. Unexpected challenges arising from the coronavirus pandemic and project deadlines prevented us from performing this experiment to our standards. We are currently working with collaborators to continue the development of the 3PWFS. A future experiment with the 3PWFS would be to analyze on a testbed the performance of the wavefront sensor at different light levels. 

\section{Conclusion}

We have presented the design of the Comprehensive Adaptive Optics and Coronagraph Test Instrument (CACTI), a new ExAO testbed. CACTI was designed with the flexibility to support visiting instruments and to be easily re-configurable to perform multiple experiments. In the current configuration, a visiting 3PWFS was integrated into CACTI for a performance test. Non-common path error was minimized by ensuring each pyramid optic had the same PSF on the tip. We demonstrated the operation of the 3PWFS by closing the AO loop on simulated turbulence on CACTI. The performance of each wavefront sensor was determined by measuring the relative Strehl ratio of the closed-loop PSF with the reference PSF. The Strehl ratio was calculated by a Strehl calculation tool developed for CACTI in Python. A difference in performance for the Raw Intensity method for the 1.6 $\lambda/D$ and 3.25 $\lambda/D$ modulation cases was on average 0.069 Strehl for the 3PWFS, and 0.086 Strehl for the 4PWFS. We cannot definitively conclude that the increase in performance of the 1.6 $\lambda/D$ modulation case was due to the increase in PWFS sensitivity, due to our systematic errors. These results are a preliminary step in the development of the 3PWFS. Due to the limitations of our system and the limited scope of our experiments we cannot definitely conclude that the Strehl values presented represent the best quantitative performance of both the 3PWFS and 4PWFS. From our results we have shown that the 3PWFS is able to reconstruct a wavefront with an accuracy comparable to the 4PWFS on the CACTI testbed when calibrated and optimized in the same way. We also found that modal loop gains must be tuned as an optimization step to maximize the performance of an ExAO system.

\section*{Acknowledgements}

This work has been supported in part by the Air Force Research Laboratory, Directed Energy Directorate, under contract FA9451-19-C-0581. The opinions, findings, and conclusions expressed in this paper are those of the authors and do not necessarily reflect those of the United States Air Force. This work has benefited from support by the NSF MRI program (AST $\#$1625441, MagAO-X) and the NASA TDEM ($\#$80NSSC19K0121).

%%%%% References %%%%%

\bibliography{article}   % bibliography data in report.bib
\bibliographystyle{spiejour}   % makes bibtex use spiejour.bst

%%%%% Biographies of authors %%%%%

\vspace{2ex}\noindent\textbf{Lauren Schatz}  is a graduate of the University of Arizona's James C.Wyant College of Optical Science. Her research focuses on developing wavefront sensing instrumentation for high-contrast adaptive optics systems. She is a 2018 ARCS Foundation scholar and a receiver of the Society of Women Engineers Ada I. Pressman Memorial Scholarship from 2019 to 2021. 

% \vspace{2ex}\noindent\textbf{First Author} is an assistant professor at the University of Optical Engineering. He received his BS and MS degrees in physics from the University of Optics in 1985 and 1987, respectively, and his PhD degree in optics from the Institute of Technology in 1991.  He is the author of more than 50 journal papers and has written three book chapters. His current research interests include optical interconnects, holography, and optoelectronic systems. He is a member of SPIE.

% \vspace{1ex}
% \noindent Biographies and photographs of the other authors are not available.

\listoffigures
\listoftables

\end{spacing}
\end{document}